\input{aipcheck}

\documentclass[
    ,final            
  ]
  {aipproc}

\layoutstyle{6x9}

\begin{document}

\title{New 6-cm observations of a large sample \\ of radio SNRs}


\author{Sebastian Soberski}{
  address={Torun Centre for Astronomy, Nicholas Copernicus University,\\ ul.Gagarina 11, 87-100 Torun, Poland}
}

\author{Eugeniusz Pazderski}{
}

\author{Andrzej Kus}{
}

\begin{abstract}
 Almost 50 radio SNR objects were selected from Green's catalogue for continuum and polarization observations . We present preliminary results of this observational campaign at 4.7 GHz carried out with Torun 32-metre radio telescope.
\end{abstract}

\maketitle


\section{Introduction}

	We present 4.7 GHz polarimetric measurements of a large sample of galactic SNRs observed using 32-metre radio telescope in Torun. About 50 objects were selected for our studies.  Supernova (SN) rates depend on star formation and are therefore related to the type of galaxy \cite{Capellaro1999}. The SNe rate in Galaxy is expected to be on the level of one SNe event in the range of 30 to 50 years. Within the last 2000 years only about 8 are identified. There are several selection effects which make SNR catalogues incomplete. For old SNRs which are expected to be intrinsically fainter than young ones, the fraction of undetected SNRs may be similar or even larger. A SN explosion within very dense material reduce the lifetime of the SNR largely. Otherwise the SNR must reach  a very large diameter, when exploding in a hot and thin environment, until it collect enough mass to become visible. These faint shell-type objects are difficult to identify in the Galactic plane because  of the superposition of a large number of emission structures along the line of sight. Outside of the Galactic plane  confusion by extragalactic sources limits the sensitivity of single-dish surveys (needed to identify extended faint objects)\nocite{Reich2002}.

\section{Observations}

\subsection{Receiving System}
Polarimetric observations were performed with Torun 32-metre radio telescope equiped with 4.7 GHz receiving system connected to polarimeter. The polarimeter was tested by observations of 3c286 at various paralactic angles and it shows that the instrumental polarization is less than 2\% \cite{Rys2006}. Some details about parameters of the receiving system are shown in Table 1.

\begin{table}
\begin{tabular}{lr}
\hline
{\bf Parametar}&{\bf Value}\\
\hline
Frequency center :& 4,7 GHz\\
Bandwidth :& 500 MHz\\
Time of integration per point :& 1 sec\\
System temperature :& 22 K\\
RMS noise (typical) :& 6 mJy/ba\\
Tb/S :& 0.2 K/Jy\\
HPBW :& 7.'2\\
\hline
\end{tabular}
\caption{Parameters of the receiving system at 32-metre radio telescope in Torun:}
\label{tab:a}
\end{table}

\subsection{First results}
 
About 70\% of 50 seleted galactic radio SNRs were observed. We present four example maps, three of them shows (Jy) polarization intensity with total power contours superimposed (Fig.1,2,3) and one displays (Jy) polarization intensity with polarization intensity contours superimposed (Fig.4). Description for maps:
\begin{itemize}
\item SNR G82.2+5.3 (W63) (Fig.1 top-left). SNR short description \cite{Green2006}: size 95x65 arcmin, type S, flux density at 1 GHz: 120? Jy, spectral index: 0.5?. Object is located in the Cygnus X complex. Spectra indicate SNR filaments at optical wavelengths. 
\item SNR G54.4-0.3 (HC40) (Fig.3 top-right). SNR short description \cite{Green2006}: size 40 arcmin, type S, flux density at 1 GHz: 28 Jy, spectral index: 0.5 . This shell is located in a complex region. Faint filaments are visible at optical wavelengths.  
\item SNR G39.7-2.0 (W50, SS433) (Fig.2 bottom-left). SNR short description\cite{Green2006}: size 120x60 arcmin, type ?, flux density at 1 GHz: 85 Jy, spectral index: 0.7?. Elongated shell, containing SS433, adjacent to the HII region S74 (radio). Faint filaments at the edge of the radio emission (optical). Emission from SS433 and two lobes (X-ray). The centre of W50 contains compact source SS433. Estimated distance is about 3 kpc (association with HI).
\item SNR G182.4+4.3 (Fig.4 bottom-right). SNR short description \cite{Green2006}: size 50 arcmin, type S, flux density at 1 GHz: 1.2 Jy, spectral index: 0.4 . Incomplete shell.

\end{itemize}

\begin{figure}
  \includegraphics[scale=0.45]{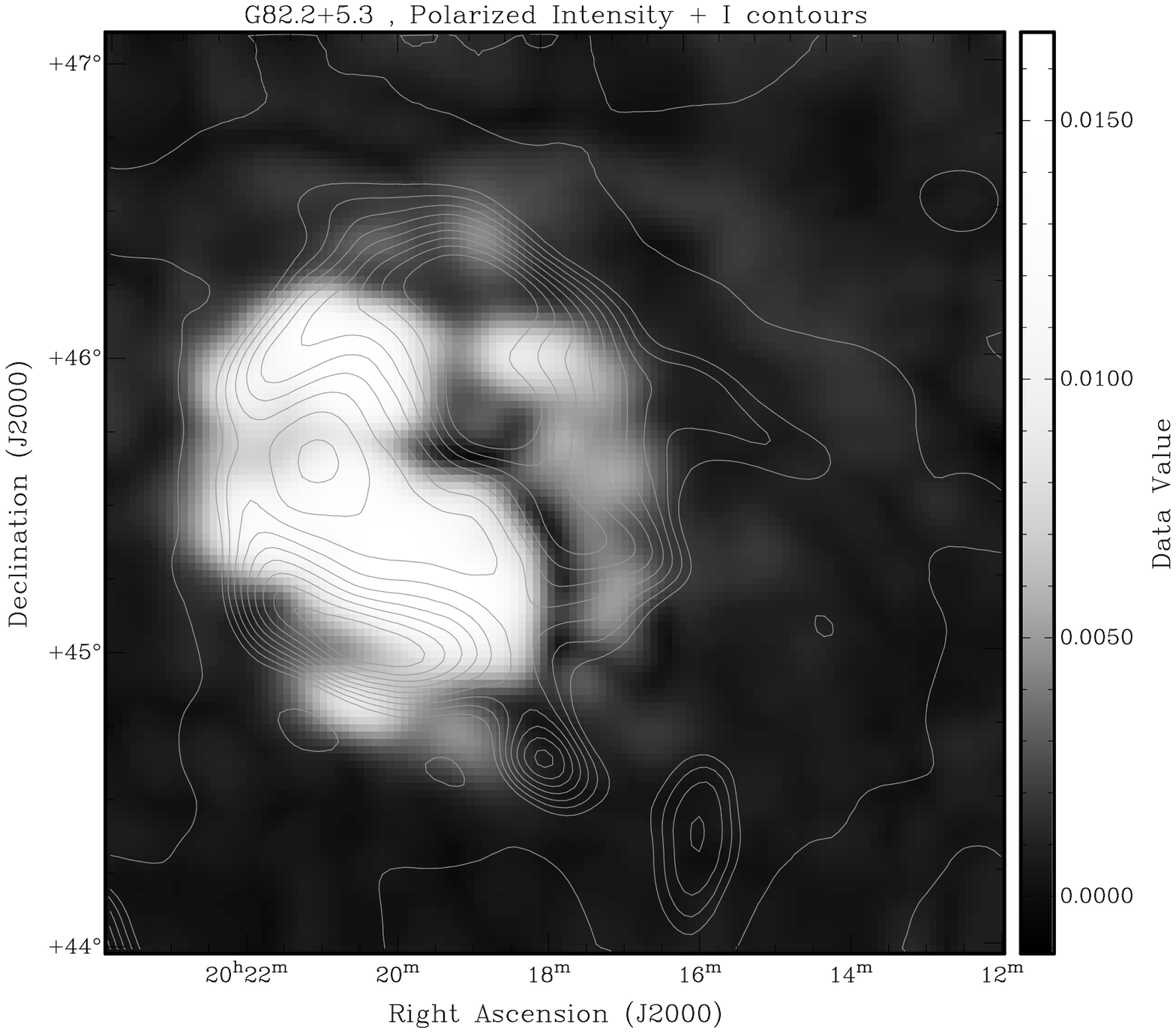}
  \includegraphics[scale=0.44]{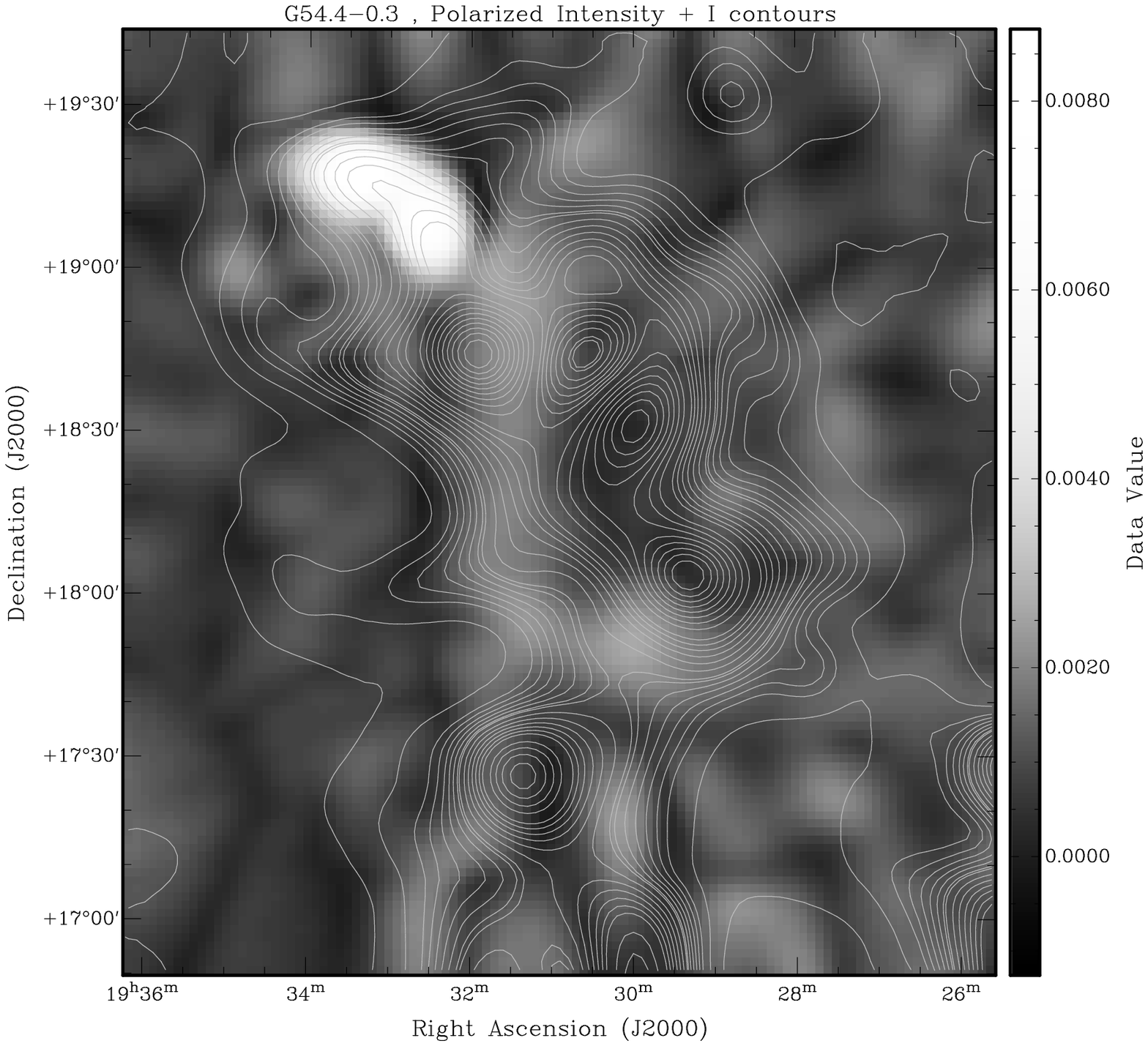}
\end{figure}

\begin{figure}
  \includegraphics[scale=0.47]{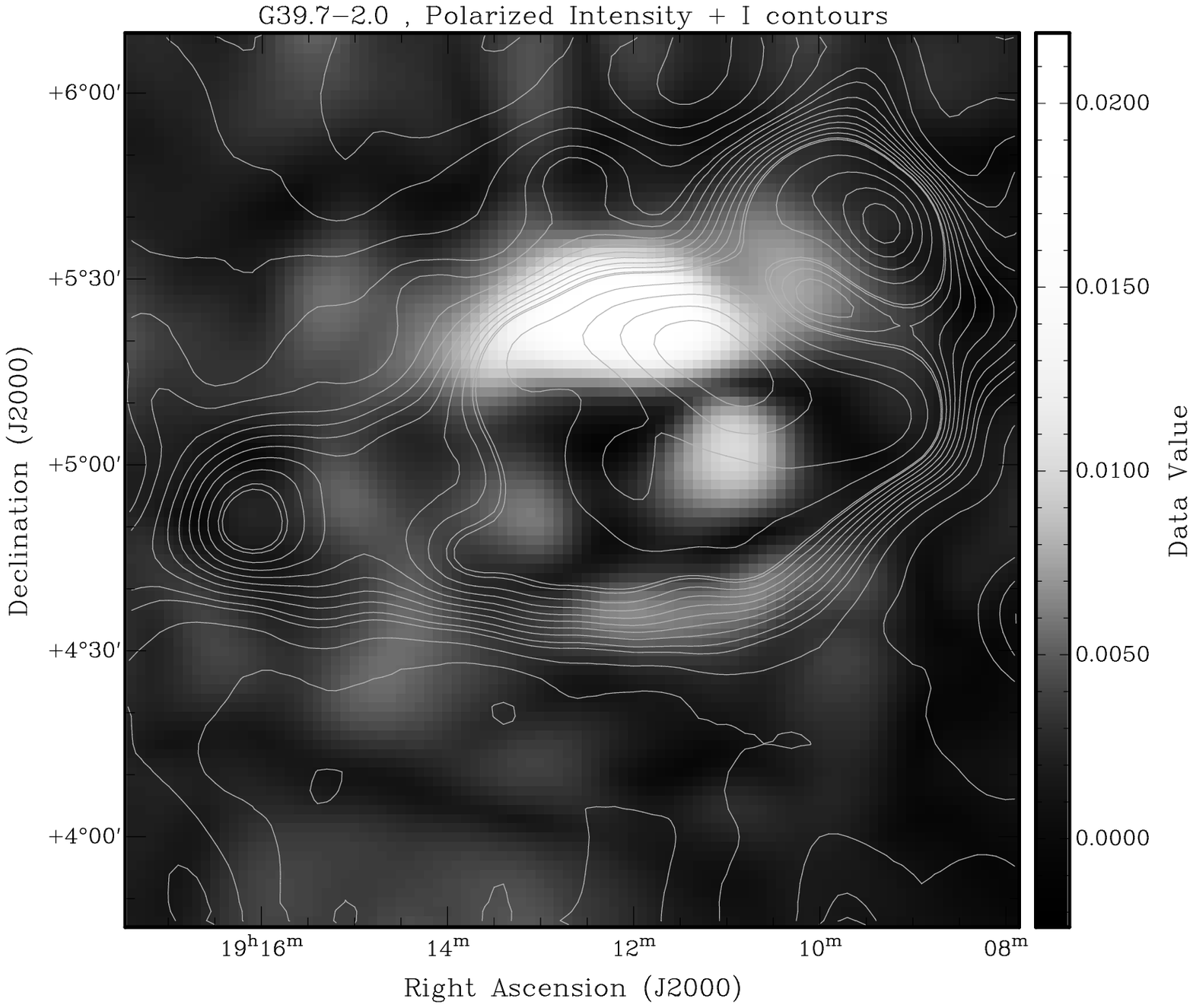}
  \includegraphics[scale=0.43]{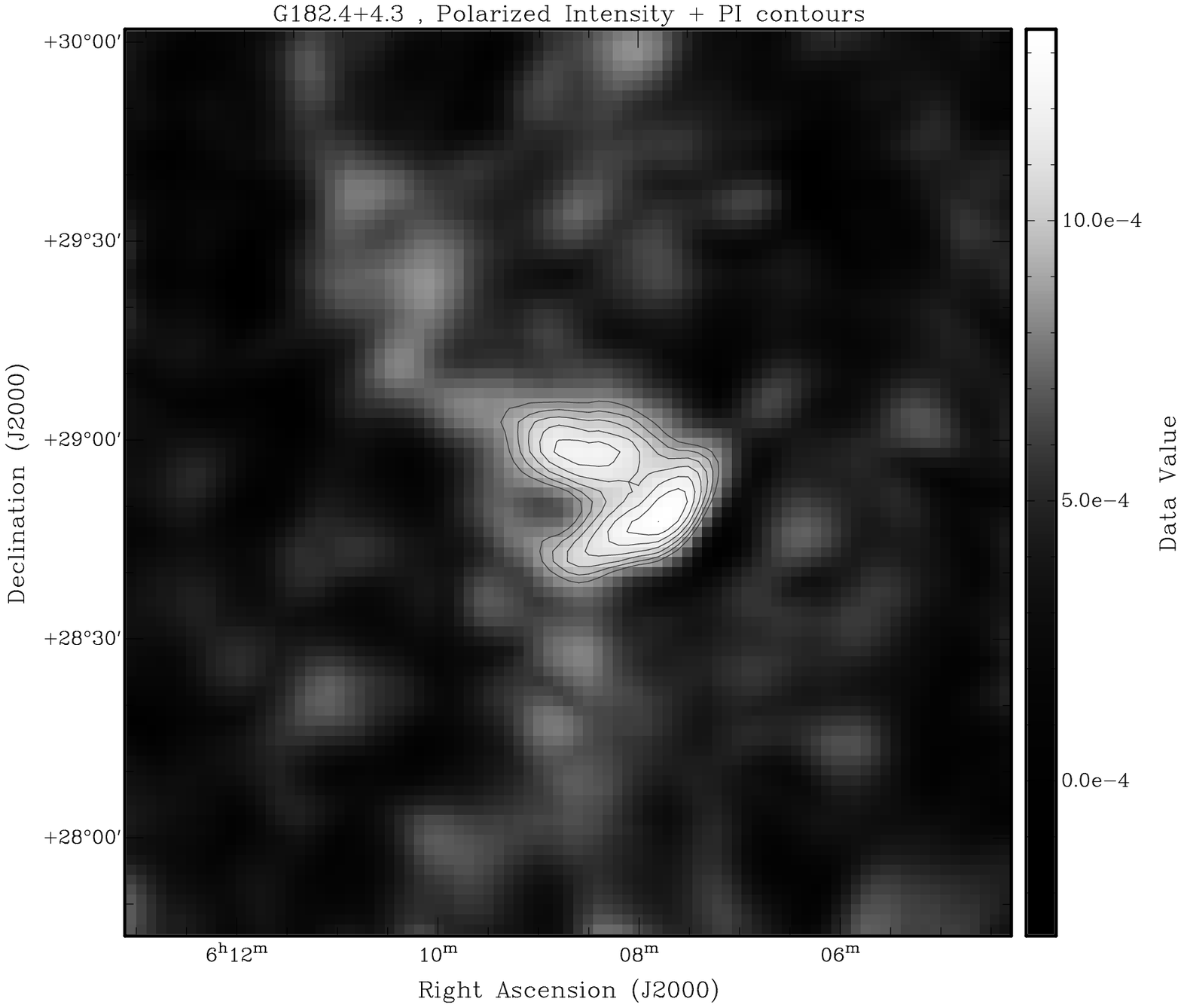}
  
\end{figure}


\begin{theacknowledgments}
  We are grateful to Prof. Richard Wielebinski and MPIfR in Bonn for many advices and the polarimeter device. We also would like to thank our colleagues from Jagiellonian University Astronomical Observatory who helped us in data reduction and analysis process.
\end{theacknowledgments}



\bibliographystyle{aipproc}   

\bibliography{soberski-aspen-bib}

\hyphenation{Post-Script Sprin-ger}
\begin{thebibliography}{4}
\expandafter\ifx\csname natexlab\endcsname\relax\def\natexlab#1{#1}\fi
\providecommand{\enquote}[1]{``#1''}
\expandafter\ifx\csname url\endcsname\relax
  \def\url#1{\texttt{#1}}\fi
\expandafter\ifx\csname urlprefix\endcsname\relax\def\urlprefix{URL }\fi
\providecommand{\eprint}[2][]{\url{#2}}

\bibitem[Cappellaro et~al.(1999)]{Capellaro1999}
E.~Cappellaro, R.~Evans, and M.~Turatto, \emph{A\&A} \textbf{351}, 459--466
  (1999).

\bibitem[Reich(2002)]{Reich2002}
W.~Reich, \enquote{{Radio Observations of Supernova Remnants},} in
  \emph{{Proc.of the 270. WE-Heraeus Seminar on Neutron Stars, Pulsars, and
  Supernova Remnants. MPE Report 278}}, edited by W.~Becker, H.~Lesch, and
  J.~Tr\"{u}mper, MPIfEP, Garching bei M\"{u}nchen, 2002, pp. 1--12.

\bibitem[Rys et~al.(2006)]{Rys2006}
S.~Rys, K.~Chyzy, A.~Kus, E.~Pazderski, M.~Soida, and M.~Urbanik, \emph{AN}
  \textbf{327}, 493 (2006).

\bibitem[Green(2006)]{Green2006}
D.~A. Green, \enquote{{A Catalogue of Galactic Supernova Remnants (2006 April
  version)},} Cavendish Laboratory, Cambridge, UK, 2006,
  \urlprefix\url{http://www.mrao.cam.ac.uk/surveys/snrs/}.

\end{thebibliography}

\IfFileExists{\jobname.bbl}{}
 {\typeout{}
  \typeout{******************************************}
  \typeout{** Please run "bibtex \jobname" to optain}
  \typeout{** the bibliography and then re-run LaTeX}
  \typeout{** twice to fix the references!}
  \typeout{******************************************}
  \typeout{}
 }

\end{document}